\documentclass[%
 reprint,%
 amsmath,amssymb,%
 aps,%
]{revtex4-2}

\usepackage{graphicx}
\usepackage{dcolumn}
\usepackage{bm}

\usepackage{microtype}

\usepackage{amsfonts}
\usepackage{latexsym}
\usepackage{dsfont}
\usepackage{amsmath}
\usepackage{amssymb}
\usepackage{amssymb}
\usepackage{amsthm}
\usepackage{xcolor}
\usepackage{setspace}
\usepackage{accents}
\usepackage{mathrsfs}
\usepackage{setspace}
\usepackage{geometry}

\allowdisplaybreaks

\def\dd{\text{d}}

\begin{document}


\title{Non-Relativistic Holography from AdS$_5$/CFT$_4$}

\author{\bf{Andrea Fontanella}}
\email[E-mail: ]{\tt andrea.fontanella[at]tcd.ie} 
\affiliation{\vspace{2mm} School of Mathematics $\&$ Hamilton Mathematics Institute, 
Trinity College Dublin, Ireland}

\author{\bf{Juan Miguel Nieto Garc\'ia}}
\email[E-mail: ]{\tt juan.miguel.nieto.garcia[at]desy.de} 
\affiliation{\vspace{2mm} II. Institut für Theoretische Physik, Universität Hamburg,
Luruper Chaussee 149, 22761 Hamburg, Germany}

\date{\today}

\begin{abstract}

We show that a novel holographic correspondence appears after taking a suitable stringy non-relativistic limit of the AdS$_5$/CFT$_4$ duality. This correspondence relates string theory in String Newton-Cartan AdS$_5\times$S$^5$ with Galilean Yang-Mills with five interacting adjoint scalars defined on the Penrose conformal boundary. As a first test, we match the Killing vectors on the string theory side with the symmetries of the dual field theory.
\end{abstract}

\maketitle


\begin{figure*}[t!]
    \centering
    \includegraphics[keepaspectratio,width=\textwidth]{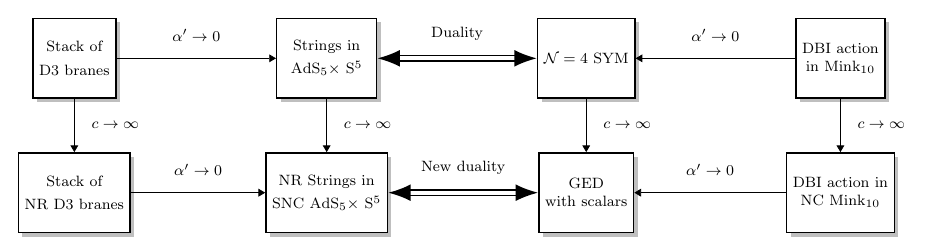}

  \caption{The top line of the diagram is the well-known relativistic AdS$_5$/CFT$_4$ correspondence. The bottom line is the new correspondence between non-relativistic theories found in this letter. In the construction, the near-horizon/decoupling limit $\alpha'\to 0$ commutes with the non-relativistic limit $c\to \infty$. }
    \label{fig:NR_holography}
\end{figure*}

\section*{Introduction}

\textcolor{red}{\underline{Correction}: In this article one should replace GED with GYM in every holographic statement, see the `Note Added' at the end of section~\ref{sec:symm}.}

\vspace{3mm}
The holographic principle states the equivalence between two seemingly unrelated theories: one describes a theory of gravity and the other a gauge field theory without gravity. The first realisation was proposed in Maldacena's celebrated work \cite{Maldacena:1997re}, and later refined in \cite{Gubser:1998bc, Witten:1998qj, Maldacena:1998im}. In his proposal, Maldacena introduced a scenario involving Type IIB string theory in flat spacetime with a stack of D3-branes. This setup allows for two distinct descriptions: one applicable at weak string coupling, featuring interactions of open and closed strings on the D3-branes, and another valid at strong string coupling, where only closed strings propagate around black D3-branes, curving the surrounding spacetime. The core idea of holography is that these two descriptions should be the equivalent at low energies, leading to the equivalence between $\mathcal{N}=4$ Super Yang-Mills (SYM) and the supergravity theory of fluctuations around the AdS$_5\times$S$^5$ geometry.

In this letter, we describe how to incorporate the non-relativistic limit into Maldacena's construction of the AdS/CFT correspondence. In principle, taking the non-relativistic limit may interfere with taking the low-energy limit, as there is no obvious reason for them to commute. For example, the near-horizon region that gives rise to the AdS$_5\times$S$^5$ geometry might not exist if the non-relativistic limit is taking first. This is what happens if we take a flat space limit of the black brane geometry. However, as we will see, this is not the case for the non-relativistic limit where the notions of horizon and conformal boundary still survive in this setting. Exploring this non-relativistic limit not only expands our understanding of holography to non-AdS spacetimes but also to non-Lorentzian geometries.

The study of non-relativistic string theory started in flat spacetime \cite{Gomis:2000bd, Danielsson:2000gi} and in AdS$_5\times$S$^5$ \cite{Gomis:2005pg}. However, it was only later discovered that the background probed by the non-relativistic string is a String Newton-Cartan (SNC) geometry \cite{Andringa:2012uz, Harmark:2017rpg, Bergshoeff:2018yvt, Bergshoeff:2019pij}. The non-relativistic string studied in these articles keeps the world-sheet relativistic, and Weyl anomalies cancel provided the beta function vanishes \cite{Gomis:2019zyu,Gallegos:2019icg}. Integrability and spectrum related problems for non-relativistic strings in SNC AdS$_5\times$S$^5$ have been studied in \cite{Fontanella:2022fjd, Fontanella:2022pbm, Fontanella:2022wfj, Fontanella:2021hcb, Fontanella:2021btt, Fontanella:2023men, deLeeuw:2024uaq}. For a review on aspects of non-relativistic string theory, see \cite{Oling:2022fft}.

On the other side, non-relativistic QFTs are relevant in many contexts of condensed matter systems. One of the simplest examples of non-relativistic QFT is Galilean Electrodynamics (GED) \cite{Santos:2004pq}, initially proposed as a Lagrangian description of the non-relativistic Maxwell equations. For a recent review on modern techniques on non-relativistic QFTs, see \cite{Baiguera:2023fus}.

In this letter, we describe how the non-relativistic limit can be incorporated into Maldacena's argument without interfering with the low-energy limit. As a result of this limit procedure on both gravity and gauge sides, we suggest a new holographic duality between non-relativistic string theory in SNC AdS$_5\times$S$^5$ and Galilean Electrodynamics in 3+1 dimensions with five uncharged massless scalar fields. This is the first example of holography involving strings with relativistic world-sheet propagating in a String Newton-Cartan background \footnote{For an example of non-relativistic holography with non-relativistic world-sheet, see \cite{Harmark:2006ta, Harmark:2016cjq}.}. As a check of our proposal, we show that there is a one-to-one correspondence between the symmetries of these theories. Our findings are summarised in Fig. \ref{fig:NR_holography}. This letter is the summary of our longer paper \cite{Fontanella:2024rvn}.

\section{The gravity perspective}

Our starting point is to consider the spacetime metric of a stack of $N$ black D3-branes, 
\begin{eqnarray}
\label{D3_metric_z}
\notag
    \dd s^2_{\text{D3-brane}} &=& \frac{4 \pi  g_s N}{\sqrt{f(z)}} \left( -\dd t^2 + \dd x^i \dd x_i   \right)  \\
    &+& \alpha'^2 \sqrt{f(z)} \left(\frac{\dd z^2}{z^4} + \frac{1}{z^2} \dd \Omega^2_5 \right) \, , \\
\notag
f (z) &=& 1 + \frac{4 \pi  g_s N}{\alpha'^2} z^4 \, ,
\end{eqnarray}
where $g_s$ is the string coupling, $(t, x^i)$, $i=1,2,3$, are coordinates along the world-volume of the D3-brane, and $\dd \Omega_5^2$ is the metric of the unit 5-sphere. We describe the 5-sphere metric in terms of Cartesian coordinates $(\phi, y^m)$, $m=1,..., 4$ and $y^2 \equiv y^m y^n \delta_{mn}$, given by 
\begin{eqnarray}
   \dd \Omega_5^2 = \left(\frac{4-y^2}{4+y^2}\right) \dd \phi^2 + \frac{4\ \dd y^2 }{\left( 4+y^2\right)^2} \, . 
\end{eqnarray}
The question we ask is whether Maldacena's near-horizon limit and the non-relativistic limit applied to the metric (\ref{D3_metric_z}) commute and give a consistent unique answer that can be trusted for non-relativistic holography. As we shall see, the answer is positive. 

\subsection{Near-horizon first, non-relativistic second.}

Maldacena's near-horizon limit of the metric (\ref{D3_metric_z}) is defined by taking $\alpha' \to 0$, giving the famous AdS$_5\times$S$^5$ metric,
\begin{eqnarray}
\label{AdS5xS5_metric}
\dd s^2_{\text{AdS}_5\times\text{S}^5} &=& R^2 \bigg(\frac{-\dd t^2 +  \dd z^2 + \dd x^i \dd x_i}{z^2} + \dd \Omega_5^2 \bigg)\, , \quad\  
\end{eqnarray}
where AdS$_5$ and S$^5$ both appear with the same radius $R^2 \equiv \sqrt{4 \pi  g_s N} \alpha'$.

As a next step, we take the non-relativistic limit on (\ref{AdS5xS5_metric}). The limit we consider is the so-called ``stringy'' non-relativistic limit, which consists in rescaling two coordinates - one time-like and one space-like - with a dimensionless parameter $c$ that ultimately will be taken to be large. The reasoning of taking the stringy non-relativistic limit instead of the ``particle'' limit, where only a time-like direction is rescaled, is because the latter case leads to a theory of non-vibrating strings \cite{Batlle:2016iel}. Instead, the stringy limit retains a non-trivial dynamics, see also \cite{Fontanella:2021btt}.   

The stringy non-relativistic limit applied to AdS$_5\times$S$^5$ was first proposed in \cite{Gomis:2005pg} in a different set of coordinates. In our set of coordinates, it translates to the following rescaling
\begin{eqnarray}
\label{NR_rescaling}
    R \to c  R \, , \ \ 
    x^i \to \frac{x^i}{c} \, , \  \ 
    \phi \to \frac{\phi}{c} \, , \ \ 
    y^m \to \frac{y^m}{c}  . \quad
\end{eqnarray}
In the $c\to \infty$ limit, (\ref{AdS5xS5_metric}) reduces to 
\begin{eqnarray}
\label{SNC_AdS5xS5_metric}
    \dd s^2_{\text{SNC AdS}_5\times\text{S}^5} &=& (c^2 \tau_{\mu\nu} + h_{\mu\nu}) \dd X^{\mu} \dd X^{\nu}\,  , \\
    \notag
    \tau_{\mu\nu} \dd X^{\mu} \dd X^{\nu} &=& \frac{R^2}{z^2} \left( -\dd t^2 + \dd z^2 \right) \, , \\
    \notag
    h_{\mu\nu} \dd X^{\mu} \dd X^{\nu} &=& \frac{R^2}{z^2} \dd x^i \dd x_i + R^2 \dd x^{i'} \dd x_{i'} \, , 
\end{eqnarray}
which is the ``String Newton-Cartan'' (SNC) version of the AdS$_5\times$S$^5$ metric \footnote{Closed strings in the background geometry (\ref{SNC_AdS5xS5_metric}) requires to couple the string to a critical closed Kalb-Ramond B-field in order to cancel the $c^2$ divergent term, as shown in \cite{Gomis:2005pg}.}. The metric $\tau_{\mu\nu}$ describes an AdS$_2$ spacetime, whereas $h_{\mu\nu}$ describes a warped Euclidean space, $w(z) \mathbb{R}^3 \times \mathbb{R}^5$, where $w(z) = z^{-2}$. Here, we denoted collectively the flat coordinates originating from the 5-sphere by $x^{i'} \equiv (\phi, y^m)$, $i'=5,...,9$, and the spacetime coordinates by $X^0 \equiv t,  X^i \equiv x^i , X^4 \equiv z , X^{i'} \equiv x^{i'}$.

Similarly to the Lorentzian AdS$_5\times$S$^5$, the geometry (\ref{SNC_AdS5xS5_metric}) also has a conformal boundary that can be described via the non-Lorentzian version of the Penrose's formalism, see e.g. \cite{Hartong:2013cba}. For that, we rescale the SNC tensors $\tau_{\mu\nu}$ and $h_{\mu\nu}$ by a conformal factor $\Omega^2 = z^2$. Then the conformal boundary is located at $z=0$, given by 
\begin{eqnarray}
\notag
    \tilde{\tau}_{\mu\nu} \dd X^{\mu} \dd X^{\nu} &=& - \dd t^2 \, , \\
    \tilde{h} _{\mu\nu} \dd X^{\mu} \dd X^{\nu} &=& \dd x^i \dd x^j \delta_{ij} \, ,  \label{NCMinkowski}
\end{eqnarray}
which describes the Newton-Cartan geometry of 4d Minkowski spacetime, NC Mink$_4$.

\subsection{Non-relativistic first, near-horizon second}

Now we want to reverse the order of taking limits, and show that the final result still remains the same. The starting point is again given by the metric of a stack of $N$ black D3-branes (\ref{D3_metric_z}). Then we implement the stringy non-relativistic rescaling given in (\ref{NR_rescaling}), supplemented with $\alpha' \to c  \alpha'$, which in the language of \cite{Avila:2023aey} is the ``transverse black brane'' limit, and shown to lead to a well-defined SNC geometry. We take $c\to \infty$, and (\ref{D3_metric_z}) becomes
\begin{eqnarray}
\label{NR_D3_metric}
    \dd s^2_{\text{NR D3-brane}} &=& \left(c^2 \tau_{\mu\nu} + h_{\mu\nu} \right) \dd X^{\mu} \dd X^{\nu}  \,  , \\
    \notag
    \tau_{\mu\nu} \dd X^{\mu} \dd X^{\nu} &=& -\frac{R^4}{\alpha^{\prime 2}} \frac{1}{\sqrt{f(z)}} \dd t^2 +  \alpha^{\prime 2}\sqrt{f(z)} \, \frac{\dd z^2}{z^4} \, , \\
    h_{\mu\nu} \dd X^{\mu} \dd X^{\nu} &=& \frac{R^4}{\alpha^{\prime 2}} \frac{\dd x^i \dd x_i}{\sqrt{f(z)}} + \frac{\alpha^{\prime 2} \sqrt{f(z)}}{z^2} \dd x^{i'} \dd x_{i'} \, . \notag
\end{eqnarray}
This is the String Newton-Cartan version of the stack of black D3-brane metric, and it retains the notion of a near-horizon region. The near-horizon region can be decoupled from the asymptotic geometry by taking the near-horizon limit $\alpha' \to 0$. By doing that, the metric (\ref{NR_D3_metric}) becomes precisely (\ref{SNC_AdS5xS5_metric}). This shows that the near-horizon and the stringy non-relativistic limits commute, giving a unique result.

\section{The gauge perspective}

Now we turn into the description of the stack of D3-branes valid at weak string coupling. The physics is governed by open and closed strings. The action is given by 
\begin{eqnarray}
\label{S_initial}
    S = S_{\text{closed}} + S_{\text{open}} + S_{\text{int}} \, ,
\end{eqnarray}
where $S_{\text{closed}}$, $S_{\text{open}}$ are the actions of closed and open strings respectively, and $S_{\text{int}}$ describes the interaction between them. 

\subsection{Non-relativistic first, decoupling second}

The starting point, as we are interested in the dynamics of $N$ coincident D3-branes in 10-dimensional Minkowski spacetime, $S_{\text{open}} + S_{\text{int}}$ is described by the non-abelian DBI action given in \cite{Tseytlin:1997csa,Myers:1999ps}. We implement the stringy non-relativistic limit by rescaling the Minkowski Cartesian coordinates as
\begin{eqnarray}
    X^i \to \frac{X^i}{c}  \,  , \qquad X^{i'} \to \frac{X^{i'}}{c}  \,  \label{NR_rescaling_gauge_1}  .
\end{eqnarray}
where $X^i$ are the space-like coordinates parallel to the D3-brane, and $X^{i'}$ are five out of the six transverse directions. To avoid divergencies, we need to set the Kalb-Ramond B-field to zero except of $B_{\mathtt{t}4} = 1$, and we are also forced to rescale the gauge potential \footnote{There exists also a second way to take the stringy non-relativistic limit, which consists in not rescaling the coordinates $X^i$ and $X^{i'}$ while rescaling the complementary coordinates by a factor $c$. This method does not require us to rescale the gauge potential. However, the symmetries of the theory obtained from this limit do not match the Killing vectors of the SNC AdS$_5\times$S$^5$ background, which we will discuss in the following section. It would be interesting to identify the string holographic dual of the gauge theory obtained from this alternative non-relativistic limit.} 
\begin{equation}
    A_\mathtt{t} \to \frac{A_\mathtt{t} }{c^2} \, , \quad A_i \to \frac{A_i }{c^2} \, , \label{NR_rescaling_gauge_2}
\end{equation}
where $\mathtt{t}$ is the time-like coordinate parallel to the D3-brane. Rescaling the gauge potential as in (\ref{NR_rescaling_gauge_2}) has the effect of pushing the commutators entering in covariant derivatives to higher orders in $c$ and, therefore, eliminating the non-abelian structure when $c\to \infty$. Effectively, this abelianise the gauge group from $U(N)$ to $U(1)^{N^2}$ \footnote{As the 3+1 spacetime is non-compact, we do not have to deal with subtleties related to global structures, as in the compact case.}.

Next, we take the decoupling limit $\alpha' \to 0$. In this way, the non-relativistic DBI action reduces to $N^2$ copies of Galilean Electrodynamics (GED) in 3+1 dimensions found in \cite{Santos:2004pq}, supplemented with 5 uncharged massless scalar fields, 
\begin{align}
\label{GED_action}
    \mathcal{L}_{\text{GED}} &= -\frac{1}{2\pi g_s} \sum_{I=1}^{N^2}\left( \frac{1}{2} \sum_{i'=1}^5 (\partial_i \phi^{i' I})^2  \right. \notag \\
    &\left.+ \frac{1}{4} (F^I_{ij})^2- F^I_{i\mathtt{t}} \partial_i \zeta^I -\frac{1}{2} (\partial_\mathtt{t} \zeta^I)^2 \right) \, .
\end{align}
where $\phi^{i' I}$ and $\zeta^I$ are coming from the six transverse fields rescaled by a factor of $2\pi \alpha'$ and after tracing away the generators.

\subsection{Decoupling first, non-relativistic second}

Once again, the starting point is the non-abelian DBI action given in \cite{Tseytlin:1997csa,Myers:1999ps}. In the decoupling limit, this action gives  $\mathcal{N}=4$ SYM in 4d Minkowski. However, as we are interested in performing the non-relativistic limit after the decoupling limit, we cannot fix static gauge yet, and also we have to include a Kalb-Ramond field.

Then we perform the rescaling (\ref{NR_rescaling_gauge_1}) and (\ref{NR_rescaling_gauge_2}). After fixing static gauge, the action diverges as
\begin{equation}
    \mathcal{L}_{\text{div.}}=-\frac{c^2}{4\pi g_s} (1-B_{\mathtt{t}4}^2 ) \sum_{I=1}^{N^2} \partial_i \zeta^I  \partial^i \zeta^I  \, ,
\end{equation}
which can be eliminated if we set the Kalb-Ramond field to $B_{\mathtt{t}4}=\pm 1$. Once we do so, the action we obtain is finite and equal to (\ref{GED_action}).
This proves that the non-relativistic and the decoupling limits commute also in the gauge theory perspective.

\section{A first test: matching symmetries} \label{sec:symm}

As a first test of our proposed non-relativistic holography, we show that the symmetries of both theories match.

All global symmetries of the action of strings with relativistic world-sheet propagating in (\ref{SNC_AdS5xS5_metric}) are given by solving the  SNC Killing equations described in \cite{Bidussi:2023rfs}. The detail of this analysis is given in \cite{Fontanella:2024rvn}. The most generic solution contains an infinite family of vector fields, given by
\begin{eqnarray}
\label{KV}
    H &=& \partial_t \, , \qquad 
    D = t\, \partial_t + z \, \partial_z + x^i \partial_i \, , \\
    \notag
    K &=& (t^2 + z^2) \partial_t +2 \,z\, t\, \partial_z +2 \, t \, x^i \partial_i\, , \\
            \notag
        P^{(n)}_i &=& (t-z)^n (t+nz) \partial_i \, , \ \  P^{(n)}_{i'} =  (t-z)^n  \partial_{i'} \, ,  \\
    \notag
    \tilde{P}^{(n)}_i &=& (t+z)^n (t-nz) \partial_i \, , \ \ \tilde{P}^{(n)}_{i'} =  (t+z)^n\partial_{i'} \, ,\\
    \notag
    J_{ij} &=& x^i \partial_j - x^j \partial_i \, , \qquad J_{i'j'} = x^{i'} \partial_{j'} - x^{j'} \partial_{i'} \, ,
\end{eqnarray}
where $n \in \mathbb{Z}$. The Killing vectors in (\ref{KV}) form an infinite dimensional algebra. The generators $H, D, K$ form an $\mathfrak{sl}(2,\mathbb{R})$ subalgebra associated with the isometries of AdS$_2$ described by the $\tau_{\mu\nu}$ metric. This infinite dimensional algebra cannot be found by standard contraction \cite{Bagchi:2009my, Bergshoeff:2023fil}, however it has an important holographic realisation. 

The symmetries of the equations of motion (on-shell symmetries) of the GED theory in 3+1 dimensions have been analysed in \cite{Festuccia:2016caf}. They are generated by  
\begin{eqnarray}
\label{GED_on_shell_symm}
\notag
    &&M^{(n)}_{i} = \mathtt{t}^{n+1} \partial_{i} \, , \quad
    H = \partial_{\mathtt{t}} \, , \quad
    D = \mathtt{t} \partial_{\mathtt{t}} + X^{i} \partial_{i} \, ,  \\
    &&K = \mathtt{t}^2 \partial_{\mathtt{t}} + 2 \mathtt{t} X^{i} \partial_{i} \, , \quad
    J_{ij} = X^{i}\partial_{j} - X^{j} \partial_{i} \, .
\end{eqnarray}
Once the Killing vectors (\ref{KV}) are evaluated at the Penrose boundary $z=0$, we immediately see that $H, D, K$, $P^{(n)}_i, J_{ij}$ precisely match $H, D, K, M^{(n)}_i, J_{ij}$ of (\ref{GED_on_shell_symm}), up to identifying the string coordinates $t$ and $x^i$ with the coordinates $\mathtt{t}$ and $X^{i}$ on NC Mink$_4$. Moreover, the generators $\tilde{P}^{(n)}_i$ become the same as the generators $P^{(n)}_i$, and therefore we just need to consider one copy of them. The same happens for $\tilde{P}^{(n)}_{i'}$ and $P^{(n)}_{i'}$.
From the string theory side, we note that we have extra generators $J_{i'j'}$ and $P^{(n)}_{i'}$ which we have not yet given a holographic explanation. In terms of the holographic gauge theory coordinates, they are realised as  
\begin{eqnarray}
    J_{i'j'} = \phi^{i'} \frac{\partial}{\partial \phi^{j'}} - \phi^{j'} \frac{\partial}{\partial \phi^{i'}} \, , \ \ 
    P^{(n)}_{i'} = \mathtt{t}^n \frac{\partial}{\partial \phi^{i'}} \, . \qquad
\end{eqnarray}
Therefore, we learn that they are realised as an infinite dimensional non-compact R-symmetry consisting in rotations $J_{i'j'}$ and time translations $P^{(n)}_{i'}$ of the fields $\phi^{i'}$. These internal symmetries are possible because the dual theory (\ref{GED_action}) does not contain commutators of the type $[\phi^{i'}, \phi^{j'}]$, nor time derivatives of the scalar fields.

\vspace{5mm}
\noindent {\bf Note Added (Aug 2025):} The proposed non-relativistic AdS/CFT correspondence was initially based on matching the on-shell symmetries of the GED theory computed in \cite{Festuccia:2016caf} with the Killing vectors of the non-relativistic string theory. Recently, the symmetries of the GED theory have been revisited in \cite{GED_symmetries}. It has been found that most of what were identified in \cite{Festuccia:2016caf} as on-shell symmetries are actually fully off-shell symmetries. Furthermore, it has been found in \cite{GED_symmetries} that the number of symmetries of the GED theory with five free scalars given in eq.~(\ref{GED_action}) is larger than the number of Killing vectors found for the bulk theory. For this reason, the proposed GED theory with five free scalars cannot be the holographic dual candidate.

As we discussed in footnote [32], there is a second consistent non-relativistic limit that one can take on the gauge theory side. This leads to a Galilean Yang-Mills theory with five non-abelian interacting scalars, whose Lagrangian was given in \cite{Fontanella:2024rvn}, 
\begin{align}\label{GYM_with_scalars}
    &\mathcal{L}_{\text{GYM}} = -\frac{1}{2\pi g_s } \sum_{I=1}^{N^2} \left( \frac{1}{2} \sum_{i'=1}^5 \bigg[ (D_\mathtt{a}\phi^{i' I})^2 \right. \notag \\
    &\mp \sum_{J,K=1}^{N^2} 2i f^{JK}\null_I \zeta^J \phi^{i' K} D_\mathtt{t} \phi^{i' I}\bigg] \notag \\
    &- \sum_{J,K=1}^{N^2}\sum_{i'<j'} \left( f^{JK}\null_I \phi^{i' J} \phi^{j' K} \right)^2 + \frac{1}{4} (F_\mathtt{ab})^2 \notag \\
    &\left. \pm F_\mathtt{at} D_\mathtt{a} \zeta^I -(D_\mathtt{t} \zeta^I)^2 %
    \vphantom{\left( \frac{1}{2} \sum_{i=1}^5 \bigg[ \right)} \right) \, ,
\end{align}
where the $\pm$ sign originates from an ambiguity of the critical B-field, and it can be absorbed into a redefinition of the $\zeta^I$ field. The symmetries of this theory have been carefully analysed in \cite{GED_symmetries}, and in contrast to footnote [32], they have been proven to be in a one-to-one correspondence with the Killing vectors evaluated at the Penrose boundary of the SNC metric of the bulk theory. This contrasts with our earlier intuition that flattening the 5-sphere pointed to the Abelian theory as the holographic dual. In light of recent developments, we are thus led to identify the GYM theory with five interacting scalars as the correct holographic dual.  

This note is also published in PRL as an Erratum of this article.

\section{Conclusions}

In this letter, we construct the first example of non-relativistic holography between a Galilean field theory and a theory of strings with a relativistic world-sheet propagating on a String Newton-Cartan background. Concretely, we propose a duality between non-relativistic string theory in SNC AdS$_5\times$S$^5$ and Galilean Electrodynamics in 3+1 dimensions with 5 uncharged massless scalar fields. Our result is based on showing that Maldacena's construction of the AdS$_5$/CFT$_4$ duality admits a non-relativistic limit that commutes with the decoupling/near-horizon limit. The final answer from both sides of the duality is unique, and we showed the new non-Lorentzian symmetries are holographically matched. 

This result breathes life into the new research area of non-Lorentzian holography as a mean to investigate holography in non-Anti-de Sitter spacetimes. There are also several questions that need to be answered regarding the quantitative test of the proposed duality. Important questions are including fermions in the construction, the matching of the partition functions and the matching of the string spectrum with the scaling dimension of dual operators, which are well-defined since GED in 3+1 dimensions is a conformal theory. Finding the first example of a non-relativistic black hole with asymptotic geometry SNC AdS$_5\times$S$^5$ would also be a very important question to explore, especially in view of the partition function matching at finite temperature.

\begin{acknowledgments}
We thank Eric Bergshoeff, Marius de Leeuw, Sergey Frolov, Troels Harmark, Jelle Hartong and Tristan McLoughlin for useful discussions. In particular, we thank Jelle Hartong for important discussions related to the Penrose conformal boundary and Killing vectors of a String Newton-Cartan geometry.
AF is supported by the SFI and the Royal Society under the grant number RFF$\backslash$EREF$\backslash$210373. JMNG is supported by the Deutsche Forschungsgemeinschaft (DFG, German Research Foundation) under Germany's Excellence Strategy -- EXC 2121 ``Quantum Universe'' -- 390833306.
AF thanks Lia for her permanent support. 

\end{acknowledgments}

\bibliography{Biblio}

\end{document}